\documentclass[doublespacing]{bmcart}

\usepackage[utf8]{inputenc} 

\usepackage{comment}
\usepackage{siunitx}
\usepackage{graphicx}
\usepackage{booktabs} 
\usepackage{mathabx}

\usepackage{endnotes}

\let\footnote=\endnote

\startlocaldefs
\endlocaldefs

\begin{document}

\begin{frontmatter}

\begin{fmbox}
\dochead{Research}


\title{The Canary in the City: Indicator Groups as Predictors of Urban Change}


\author[
   addressref={aff1,aff2},                   
   corref={aff1},                       
   email={steentoft@arch.ethz.ch}   
]{\inits{AS}\fnm{Aike A} \snm{Steentoft}}
\author[
   addressref={aff3},
   email={ate_poorthuis@sutd.edu.sg}
]{\inits{AP}\fnm{Ate} \snm{Poorthuis}}
\author[
   addressref={aff2},
   email={EBSLEE@ntu.edu.sg}
]{\inits{BL}\fnm{Bu-Sung} \snm{Lee}}
\author[
   addressref={aff1,aff2},
   email={schlaepfer@arch.ethz.ch}
]{\inits{MS}\fnm{Markus} \snm{Schl{\"a}pfer}}


\address[id=aff1]{
  \orgname{Future Cities Laboratory, Singapore-ETH Centre}, 
  \cny{Singapore}                                    
}
\address[id=aff2]{%
  \orgname{Nanyang Technological University},
  \cny{Singapore}
}
\address[id=aff3]{%
  \orgname{Singapore University of Technology and Design},
  \cny{Singapore}
}



\end{fmbox}


\begin{abstractbox}

\begin{abstract} 


As cities grow, certain neighborhoods experience a particularly high demand for housing, resulting in escalating rents. Despite far-reaching socioeconomic consequences, it remains difficult to predict when and where urban neighborhoods will face such changes. To tackle this challenge, we adapt the concept of `bioindicators', borrowed from ecology, to the urban context. The objective is to use an `indicator group' of people to assess the quality of a complex environment and its changes over time. Specifically, we analyze 92 million geolocated Twitter records across five US cities, allowing us to derive socio-economic user profiles based on individual movement patterns. As a proof-of-concept, we define users with a `high-income-profile' as an indicator group and show that their visitation patterns are a suitable indicator for expected future rent increases in different neighborhoods. The concept of indicator groups highlights the potential of closely monitoring only a specific subset of the population, rather than the population as a whole. If the indicator group is defined appropriately for the phenomenon of interest, this approach can yield early predictions while simultaneously reducing the amount of data that needs to be collected and analyzed.
\end{abstract}




\end{abstractbox}
%

\end{frontmatter}



\section*{Introduction}

\textbf{Urban change.} Cities around the world continue to attract additional residents who, in many cases, can be quite discerning about what type of neighborhood they want to live in. As a result, the housing market in many urban neighborhoods has experienced increased pressure, often leading to escalating prices and the loss of affordable housing for urban residents. However, despite far-reaching consequences, it remains difficult to predict when and where neighborhoods experience these pressures. Previous work has often evaluated the characteristics of the neighborhood itself: e.g., distance to employment centers, schools, restaurants and other amenities, crime level, and a wide array of demographic data such as income, education and age of its residents \cite{hill2008hedonic}. However, collecting such data is resource intensive, which often results in data only collected decennially or, in a best case, annually. This limits the applicability in analysis of urban processes that operate and change on a much finer temporal scale, such as the early prediction of rising housing prices in urban neighborhoods. While recent research has been increasingly looking at monitoring the city in near real-time by making use of new types of user-generated data \cite{hristova2016measuring}, the link between collecting and monitoring this type of data and its actual use for predicting future changes in urban social processes remains understudied.

\textbf{The canary in the city.} \textit{``What can the canary in the coal mine tell us? Historically, canaries accompanied coal miners deep underground. Their small lung capacity and unidirectional lung ventilation system made them more vulnerable to small concentrations of carbon monoxide and methane gas then their human companions. As late as 1986, the acute sensibility of these birds served as biological indicator of unsafe conditions in underground coal mines in the United Kingdom. [..] All species (or species assemblages) tolerate a limited range of chemical, physical, and biological conditions, which we can use to evaluate environmental quality."} \cite{holt2011bioindicators}

In this paper, we apply the concept of bioindicators, borrowed from ecology \cite{holt2011bioindicators}, to a social science context. The overall objective is to use an `indicator group' of people to assess the `quality' of a neighborhood and its change over time. Conventional approaches typically conduct social, economic and infrastructural studies \textit{on the entire population} to directly measure urban parameters. Instead, we use a much smaller \textit{subset of the population} -- the indicator group -- as a proxy for a neighborhood's condition. The starting premise is that certain subgroups within the population (e.g., young affluent singles) have a different tolerance level for certain neighborhood characteristics than others (e.g., less affluent families). Just like the canary was employed as a practical proxy for levels of carbon monoxide and methane, so the presence or absence of such a subgroup can be used as a practical proxy for difficult-to-measure neighborhood characteristics that might precede rising housing prices (e.g., the `hipness' of an area). Instead of trying to measure, for example, trendy restaurants, we monitor the people who might make use of them.

The use of indicator groups is fundamentally different from more conventional measures of neighborhood characteristics and offers a key advantage. In contrast to location-based measures such as census data, it is an individual-based measure that allows to track near real-time how people make use of urban space. This, in turn, allows to detect changing usage patterns \textit{before} they are manifested in census data. An illustrative example are neighborhoods that attract more young affluent singles before housing prices are increasing \cite{lees2010gentrification}. Thereby, seemingly small events may have a large impact on these usage patterns. For example, a few isolated crime incidents can seem insignificant in the official statistics (and might not even be easily detected), but might have a drastic effect on the neighborhood's attractiveness. As such, the specific `tolerance' of an indicator group to such small incidents can provide early warning signals of potentially strong changes in the neighborhood characteristics.

\textbf{Social sensing.} While bioindicator studies in ecology often depend on the development of systems that identify and track specific, individual species, urban studies differ from this in a significant way: human beings all belong to the same species. As such, the challenge is not selecting a pre-defined group but rather defining a specific subset, or indicator group, from within the whole population. The great advantage of the current computational landscape is that human beings, through their engagement with `smart' devices and the Internet, have become `sensors' themselves and are now continuously producing a wide variety of data about their behavior. Many of today's computing applications, social media platforms being a prime example, allow users to connect with each other and share content as well as their location, allowing the extraction of the `who' and the `where' at an unprecedented scale. In contrast to classic census or survey studies, social sensors can provide a high spatial and temporal resolution, allowing studies on different spatial scales in near real-time \cite{pentland2010honest}. All of this data allows us to define and adapt appropriate and precise indicator groups for different social phenomena.

\textbf{Present work.} In this work we develop a methodology to monitor neighborhoods based on the characteristics of its visitors, allowing us to predict neighborhoods with expected rent increase in subsequent years. We combine the bioindicator concept, borrowed from ecology, with social sensing data from location-based social networks. In the present work, we use data from Twitter, which allows us to study the spatial behavior of 1.6 million users in the US cities Chicago, New York City, Los Angeles, Boston and Portland over the course of one year (July 2012 - June 2013). During this time, these users created 92 million geotagged tweets. We set up a bipartite network of user nodes, location nodes and visitor links, allowing us to measure how often each user tweeted within each neighborhood. Based on this, we can create a profile for each user based on the socio-economic characteristics of the visited neighborhoods. Specifically, we assign income and age profiles that are then used to identify appropriate indicator groups. In the following, we refer to the users outside the 90 percent quantiles as high-income-profile and low-income-profile users and high-age-profile and low-age-profile users respectively. A given user with a high-income profile might not necessarily have a high income himself but just \textit{visits} high-income neighborhoods more often. We find that the number of visitors with a high-income (low-income) profile shows a strong positive (negative) correlation with the neighborhood's rent level. Further, we find that the number of high-income-profile visitors between July 2012 and June 2013 is strongly correlated with an increase of the rent values over the \textit{following} years (2012 - 2015). Importantly, by examining a number of neighborhoods that are currently known for their increasing rent levels and associated socio-economic changes, we find that those areas tend to have a disproportionately low rent level compared to other neighborhoods with a similar number of high-income-profile visitors. Finally, we further enhance the indicator group by including the age profile of the users. Here, we find that users with a high-income, low-age profile are especially well-suited as an indicator group: their visits to a neighborhood are an early predictor of drastic future rent increase. The adaptation of the concept of bioindicators to urban residents, monitored through social sensing data, allows for the development of a broad range of applications with relevance for various stakeholders from urban planning and governance to the real estate industry. While we focus on rising rents in neighborhoods in this paper, our methodology is completely generic and can potentially be applied to a broad range of urban challenges, such as detecting social segregation processes or predicting spatio-temporal crime patterns in cities.

\section*{Related work}

The current work builds forth on three distinct domains of related work. The first is a long-standing tradition of studying urban processes and urban change. Why and how does a city, its neighborhoods and its people, adapt and evolve? Second, and related, is the domain of social sensing that looks at these questions through the lens of the many novel data sources and computational methods that have emerged in the last twenty years. Although this new wave of data may overcome shortcomings of conventional urban research (e.g., lack of granular data), social sensing has its own set of challenges when trying to understand, predict and, ultimately, govern the dynamics of cities. It is here that we introduce the third domain, adapted from ecology, and posit the concept of bioindicators as a useful method in the practical application of social sensing.

\textbf{Urban Change and Housing.} There is a long tradition of studying urban change, specifically in relation to housing, within the social sciences \cite{zukin1989loft, glass1964london, smith1982gentrification}.  These indicators can range from median household income and educational attainment to housing prices and the number of amenities. Generally speaking, these studies try to understand why and how urban change takes place \cite{atkinson2000measuring} and which neighborhoods are currently undergoing change \cite{hammel1996model}. Another tradition has focused on understanding and modeling neighborhood change as a process, often using ecological concepts such as the invasion-succession model \cite{hoover1959anatomy, schwirian1983models}. Recently, this approach has been reinvigorated with computational methods adopted from across disciplines that allow for an empirical analysis of neighborhood trajectories through time, such as Self-Organizing Maps \cite{delmelle2017differentiating} and genetic sequencing \cite{delmelle2016mapping}. However, most studies still rely on census data or other conventional survey instruments and track socio-economic indicators aggregated to the neighborhood level.

\textbf{Social sensing.} Answering the question of \textit{`who' moves `where' and `when' in the city} is of great interest for social and economic research and policy. Novel sources of user-generated data such as those automatically collected from social media or mobile phones have proven to provide valuable new insights into the organization of cities \cite{ratti2006mobile,schlapfer2014scaling,shelton2015social, zhong2017centrality}. This geo-referenced data is unprecedented both in terms of the number of people it covers and in terms of its spatiotemporal resolution, which also enables new approaches to the demographic profiling of people. Blumenstock et al. \cite{Blumenstock2015predicting} recently demonstrated that feature engineering techniques allow for the inference of different socio-economic characteristics of individuals from their anonymized mobile phone usage patterns. Calabrese et al. \cite{calabrese2013understanding} inferred the home location of the user to then assign socio-economic census data. Other approaches go beyond only using the spatio-temporal aspects of this data and analyze other variables contained within user-generated data. For example, Facebook Likes can be used to make predictions about the age, gender, and ethnicity of the user \cite{kosinski2013private} and the content of tweets can be used to predict the occupational class of the user \cite{preoctiuc2015analysis}. By integrating multiple data sources from Foursquare, Twitter, Instagram, and Facebook, Farseev et al. \cite{farseev2015harvesting} were able to create  mobility and demographic profiles for social media users. These studies have enabled great methodological advances towards a near real-time monitoring of cities - although a widespread adoption of applications of these methods has still to emerge.

\textbf{Indicator group.} In ecology the concept of bioindicators is often employed to study the quality of environments -- and monitor potential changes. Although bioindicators may contain a variety of biological processes, a single species or a collection of species is usually taken. Not every species makes for a good bioindicator: bioindicators have a very low tolerance to changes in a particular factor of interest to the researcher; much lower than other species in the same environment \cite{holt2011bioindicators}. Different bioindicators can be identified and developed for different types of environments. An overview of the concept, including methodology, benefits and disadvantages, is given by Holt et al. \cite{holt2011bioindicators} while Siddig et al. \cite{siddig2016ecologists} provide a thorough overview of how ecologists have selected, used, and evaluated the performance of bioindicators over the last 15 years.

To the best of our knowledge, the concept of bioindicators or, more generally, indicator groups has not seen uptake within the social sciences. Perhaps the closest adaptation is within the field of geodemographics \cite{singleton2014past}. Geodemographics aims to segment the population into smaller subgroups that go beyond traditional indicators (e.g., age, education) and more towards 'lifestyle'. In commercial applications this can be used for customer targeting \cite{birkin1995customer, esri2016tapestry} but it can also be used for the better targeting of public policy \cite{petersen2011geodemographics}. Although the use and identification of these segments is not dissimilar from the construction of indicator groups, geodemographic segmentation has been exclusively used in an applied manner (e.g., for customer targeting) rather than the use of a segment as an indicator group in subsequent scientific research.

\section*{Monitoring model}

\begin{figure}[b!]
    \includegraphics[width=6cm]{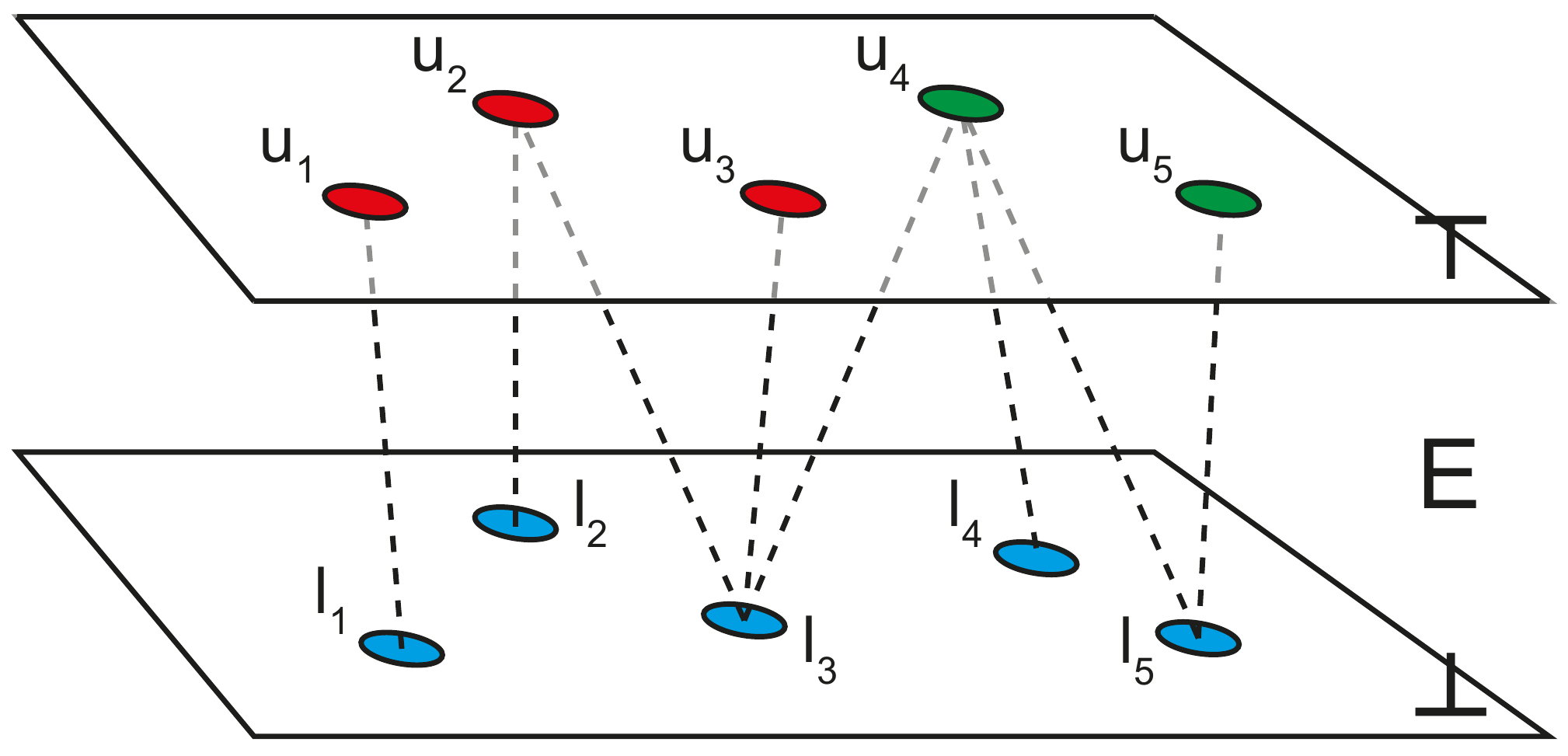}
    \caption{\csentence{Model for monitoring indicator group.} 2-mode network model that consists of a set of user nodes in the top layer (indicator users, green), a set of neighborhood nodes in the bottom layer, and a set of links representing the visits of users to neighborhoods. Adapted from \cite{hristova2016measuring}.}
    \label{fig:network}
\end{figure}

Figure \ref{fig:network} presents our monitoring model that allows to quantify the visiting frequency of an indicator group to neighborhoods. We set up a bipartite graph $G = (\top, \bot, E)$ where $\top$ is the set of top nodes, $\bot$ is the set of bottom nodes, and $E \subseteq \top \times \bot$ is the set of links. The top layer consists of a set of nodes $u \in \top$ that represents a set of $N$ users where each user $u_n$ can be described with a unique user id. The bottom layer consists of a set of nodes $l \in \bot$ that represents a set of $M$ neighborhoods where each neighborhood $l_m$ can be described by a set of polygons with vertices described by coordinates. A set of links $v \in E$ represent the visits of users to neighborhoods where $v_{n,m}$ can be described as the number of visits of user $u_n$ to neighborhood $l_m$. A visit is made when the user was spatially located in the neighborhood. For the definition of $v_{n,m}$, we use a temporal resolution of one day. Therefore, several (potentially) distinct visits of a given user to the same neighborhood during the same day (midnight - midnight) are treated as one single visit.

Each node $u_n$ has a set of K node attributes represented by column vector $A_n$ of size K and is denoted as the user profile. The user profile is evaluated based on the visited neighborhoods:
\begin{displaymath}
A_n = \sum_{m=1}^{M} w_{n,m}B_{m}
\end{displaymath}
where $B_m$ is a column vector of size K that represents a set of K node attributes of node $l_m$ and is denoted as the residents profile, and $w_{n,m} = v_{n,m} / {\sum_{m=1}^{M} v_{n,m}}$ normalizes the number of visits of user $u_n$.

This allows us to segment users based on their spatial profile and define our indicator group I as a set of L indicator users where each indicator user $u_l$ fulfills the indicator group requirements for all attributes K. Finally, we define for each neighborhood $l_m$ the visitor profile $p_m$ as the number of unique indicator group visitors in a given time period.

\section*{Dataset}

\textbf{Visitors.} The availability of new types of data collected from human beings or devices on their behalf opens up the opportunity to monitor the movement of people and therewith the `pulse' of the city at an unprecedented scale near real-time. Here, mobile devices function as sensors that can provide information about the `who', the `where' and the `when' of urban life. For this work, we collected a dataset of 92 million geotagged tweets through the Twitter streaming API \cite{poorthuis2017making}. The tweets were posted by 1.6 million users across five US cities, namely Chicago, New York City, Los Angeles, Boston and Portland (spatial boundaries as defined by the Metropolitan Statistical Areas), in the time period between July 2012 and June 2013. Each geotagged tweet contains information about the user described by an anonymized user ID, the location from which the tweet was sent as lat/lon coordinates and the time at which the message was created.

\textbf{Residents.} For socio-economic neighborhood attributes, as well as the spatial delineation of neighborhood boundaries, data from the U.S. Census Bureau is used. The data from the Census Bureau provides an extensive snapshot of the characteristics of urban residents, based on a statistically representative sample of the society, and also covers the most relevant socio-economic domains. Specifically, we employ a dataset based on the 5-Year Estimates from the American Community Survey (ACS) that contains the median household income and median age in a neighborhood for the year 2012. For the neighborhood boundaries, we use the census tract definition, which are areas with a population between 2,000 and 8,000 residents.

\textbf{Housing prices.} Collecting accurate, comparable and reliable housing price data over the period of several years on a small spatial scale such as the census tract level is a difficult task as transaction prices and rental values are often not publicly available. For this work, we again employ a dataset based on the ACS 5-Year Estimates, which contains the median contract rent on census tract level for the years 2012 and 2015. We do need to note here that a drawback of ACS data at this spatial scale is its large margins of error on census tract level, a result of a relatively small sample size. In the following, we refer to the census tracts as neighborhoods. 

\section*{Predicting urban change}

\subsection*{Selection of the indicator group}

\textbf{Challenges.} The first challenge is the selection of an appropriate indicator group for the phenomenon of interest. Not all social groups can serve as successful indicator groups for the same phenomenon and, more so, social groups are not discrete, natural units like species.
The second challenge is the derivation of a user profile based on human activity data. Here, recent research made substantial advances towards an accurate and reliable feature extraction, for example, through the inference of different socio-economic characteristics from mobile phone data \cite{Blumenstock2015predicting} or social media data such as Facebook Likes \cite{kosinski2013private}. A common difficulty is the access to fine-grained geotagged data at the individual level. For this work, we employ a large dataset of geotagged tweets, which are publicly accessible through the Twitter streaming API. However, when zoomed into the individual and neighborhood level, geotagged tweets become relatively sparse. Combined with the fact that only a small subset of Twitter users opts-in to sharing their location \cite{hecht2011tweets, poorthuis2017making}, the inference of profiles for a broad set of users, let alone an entire population, is especially challenging.

Finally, we need to monitor the indicator group's activity across the city and find an appropriate measure that can be used as a predictor for the housing prices.

\textbf{Method.} Previous studies have shown that urban change is strongly related to socio-economic attributes of the neighborhood, such as income or age profiles \cite{case1990forecasting}. These studies give us a set of relevant domains for urban change predictions but they lack a specific recipe for selecting indicator groups. 
As a proof-of-concept, we define income as our first indicator attribute and select a specific income range as our indicator feature. We select the indicator group according to the income-profile distribution for all users in the dataset. To test the accuracy of the indicator group selection, we define a reference group of similar size but with a different income range. Exemplary, and to showcase the different performances of different data subsets, we define the 10 percent `richest' users as our indicator group, denoted as high-income-profile users, and the 10 percent `poorest' users as the reference group, denoted as low-income-profile users. It is important to note that it is not our goal to provide an explanation for the underlying complex process of rising housing prices \cite{NBERw11129}, but rather to have available a simple but powerful indicator that signals potential future changes in rents before they are manifested in, for instance, census data.

To estimate an income profile, we employ a relatively simple model that characterizes users based on the characteristics of the neighborhoods they visited. This makes theoretical sense as well: for the prediction of urban change, not only the actual income of visitors to a neighborhood is relevant but also their tastes and preferences. These tastes and preferences are proxied by the character of the other neighborhoods they visit. Specifically, to each user we assign the weighted median household income of visited neighborhoods, where the weight is the number of visits to the neighborhood divided by the number of visits to all neighborhoods. We define the number of visits as the number of distinct days in one year a user was geolocated in the neighborhood. Due to the sparseness of geotagged tweets, we used the visit history of a user over the course of an entire year (July 2012 - June 2013). Further, we excluded users with less than 10 visits for this time period. 

Finally, we need a measure that reflects how frequently our indicator group is attracted to each neighborhood. To do so, we define the indicator measure as the number of unique indicator visitors in a neighborhood in a given time period. By counting the number of unique indicator visitors, rather than their total visits, we avoid an over-representation of power users. The indicator measure is independent of the time of visit, day in the week, duration of stay, and frequency of visits. It simply reflects how many unique high-income-profile users a neighborhood attracted over the course of one year.

When comparing the indicator measure to housing prices at the neighborhood level, we face the challenge that ACS data at this spatial scale has relatively large margins of error. It gets even more difficult when analyzing the housing prices over time, as the errors for the same neighborhood but for different years are independent. To resolve this issue and to make correlations visible, we use neighborhood bins instead, which consequently reduces the variability. Bin edges represent thresholds for the number of visitors and are spaced evenly on a log scale, while the bin value represents the median of all included samples within those edges. Neighborhood bins that contain less than 10 neighborhoods are removed. We denote the resulting number of neighborhood bins as $N$.

\textbf{Indicator group statistics.}
\begin{table}[t!]
    \sisetup{scientific-notation=false, round-precision=1, round-mode=places}
    \caption{Network statistics for indicator group.}
    \label{tab:monitoring_multiple}
    \begin{tabular}{lrrrrr}
        \toprule
        City&\# Users&\# Neighb.&\# Tweets&\# Visits&\# Links\\
        
        \midrule
        Chicago&\num{8238}&\num{2032}&\num{1516730}&\num{478507}&\num{118081}\\
        NYC&\num{11284}&\num{1554}&\num{1430679}&\num{469957}&\num{115696}\\
        LA&\num{13638}&\num{2730}&\num{2219601}&\num{712494}&\num{201468}\\
        Boston&\num{3706}&\num{579}&\num{605596}&\num{202319}&\num{37755}\\
        Portland&\num{1546}&\num{426}&\num{311468}&\num{83464}&\num{9893}\\
        \bottomrule
    \end{tabular}
\end{table}
Table \ref{tab:monitoring_multiple} shows the descriptive statistics of the monitoring model based on our indicator group in five US cities. Taking Chicago as an illustrative example, we identified 8,238 users as indicator users, which represents 10 percent of the entire user population. Together they visited 2,032 different neighborhoods. In the period between July 2012 and June 2013, they sent around 1.52M tweets which translates into around 0.48M visits or 0.12M network links. Hence, on average, an indicator user visited around 14 different neighborhoods, and each of those about 4 times per year.

\begin{figure}[b!]
    \includegraphics[width=0.7\linewidth]{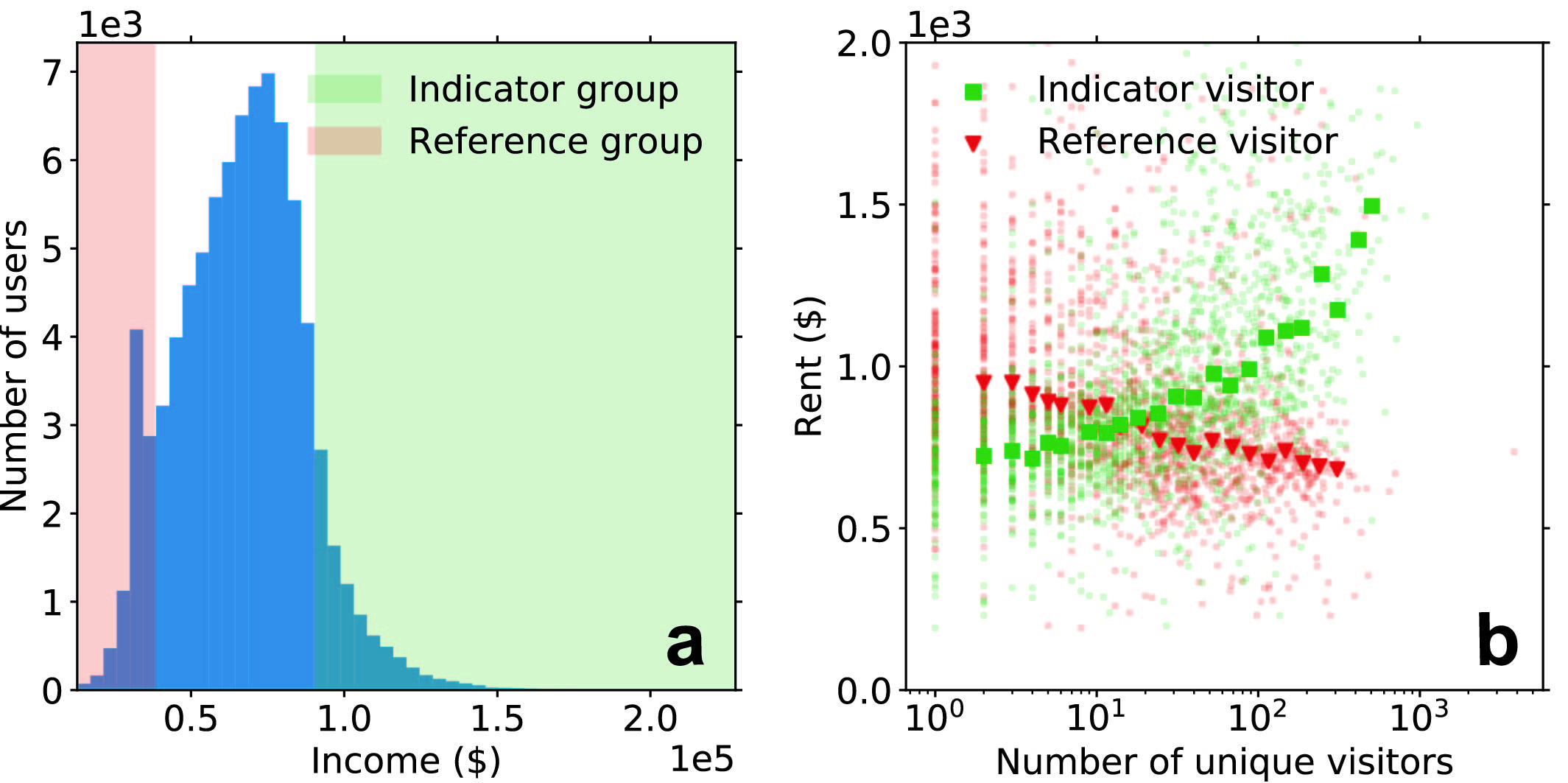}
    \caption{\csentence{Selecting an indicator group.} \textbf{a}, Income-profile distribution for users in Chicago. We assign an income profile to each user based on the census median income of the visited neighborhoods. The 10 percent `richest' users represent the indicator group (green) while the 10 percent `poorest' users function as a reference group (red). \textbf{b}, For each neighborhood, we compare the number of unique indicator visitors (green) and the number of unique reference visitors (red) in 2012/2013 with the neighborhood's median rent prices in 2012. Neighborhoods that bring together more indicator visitors show higher rent prices while more reference visitors indicate lower rent prices.}
    \label{fig:rent_Chicago}
\end{figure}

Figure \ref{fig:rent_Chicago}a shows the distribution of the derived income profile for all users in Chicago. The indicator group (green) and the reference group (red) correspond to the 10 percent `richest' and 10 percent `poorest' users, respectively. Figure \ref{fig:rent_Chicago}b confirms our initial assumption that a higher number of high-income-profile visitors is associated with higher rents, as indicated by Spearman's rank correlation coefficient ($r_s$)  ($r_s=0.99$; $p<10^{-18}$; $N=22$), and that a higher number of low-income-profile visitors is an indicator for lower rents ($r_s=-0.97$; $p<10^{-12}$; $N=20$). Interestingly, towards a higher number of unique visitors the two curves diverge the most, indicating large differences between neighborhoods with high numbers of high-income-profile visitors and neighborhoods with high numbers of low-income-profile visitors (see also Figure \ref{fig:deltarent_Chicago}b and d). This strengthens our initial assumption that defining an indicator group based on the income profile of visited location is an appropriate measure when looking at future changes in housing prices.

\subsection*{Predicting neighborhood change}

\textbf{Rising housing prices.}

\begin{figure}[t!]
    \includegraphics[width=0.65\linewidth]{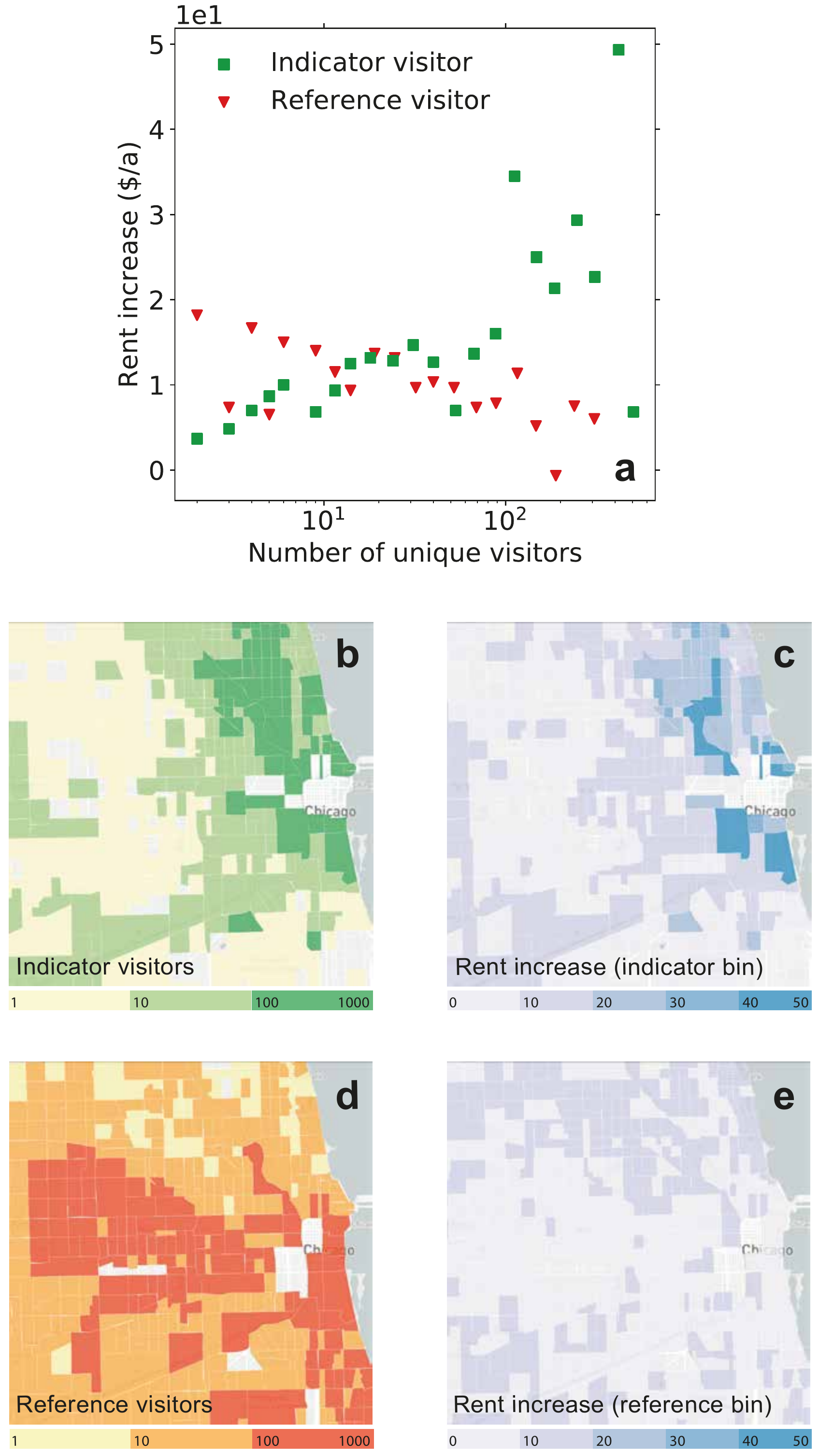}
    \caption{\csentence{Comparing the predictions of the indicator group and the reference group.} \textbf{a}, For each neighborhood bin in Chicago, we compare the number of unique indicator visitors (green) and the number of reference visitors (red) in 2012/2013 with the 2012-2015 change of the census median rent. Neighborhoods with more indicator visitors tend to experience a higher rent increase. \textbf{b-c}, Mapping the rent price predictions (a) for the indicator group, and \textbf{d-e}, for the reference group.}
    \label{fig:deltarent_Chicago}
\end{figure}

Figure~\ref{fig:deltarent_Chicago} demonstrates the specific suitability of our indicator measure for the prediction of rising rents within the different neighborhoods of Chicago. Indeed, as is depicted in Figure~\ref{fig:deltarent_Chicago}a, the number of high-income-profile visitors between July 2012 and June 2013 is strongly correlated with an increase of the rent values over the \textit{following} years (2012 - 2015) ($r_s=0.70$; $p<10^{-3}$; $N=22$). Conversely, the correlation becomes negative for our reference group ($r_s=-0.59$; $p<10^{-2}$; $N=20$). This confirms our main assumption behind the selection of our indicator group: neighborhoods that attract a large number of visitors with a high-income profile also tend to experience rising rents in the subsequent years. Figures~\ref{fig:deltarent_Chicago}b-e show the spatial aspect of the revealed relations as choropleth maps. Comparing Figure~\ref{fig:deltarent_Chicago}b with Figure~\ref{fig:deltarent_Chicago}d reveals a strong difference between the activity spaces of the indicator group and the reference group. The neighborhoods preferred by the indicator group are located towards the north of the city center, while those preferred by the reference group tend to be located in the western and southern parts of the city center. This difference is not surprising per se. It simply reflects the spatial segregation along socio-economic lines common in many cities. However, it does again underline that our indicator group is in some ways distinct from the population as a whole. Moreover, comparing Figure~\ref{fig:deltarent_Chicago}b with Figure~\ref{fig:deltarent_Chicago}c, visually confirms that neighborhoods with a large number of indicator visitors also experience a high rent increase in the following years. Comparing Figure~\ref{fig:deltarent_Chicago}d with Figure~\ref{fig:deltarent_Chicago}e indicates the negative correlation found for the reference group.
\begin{figure}[t!]
    \includegraphics[width=0.75\linewidth]{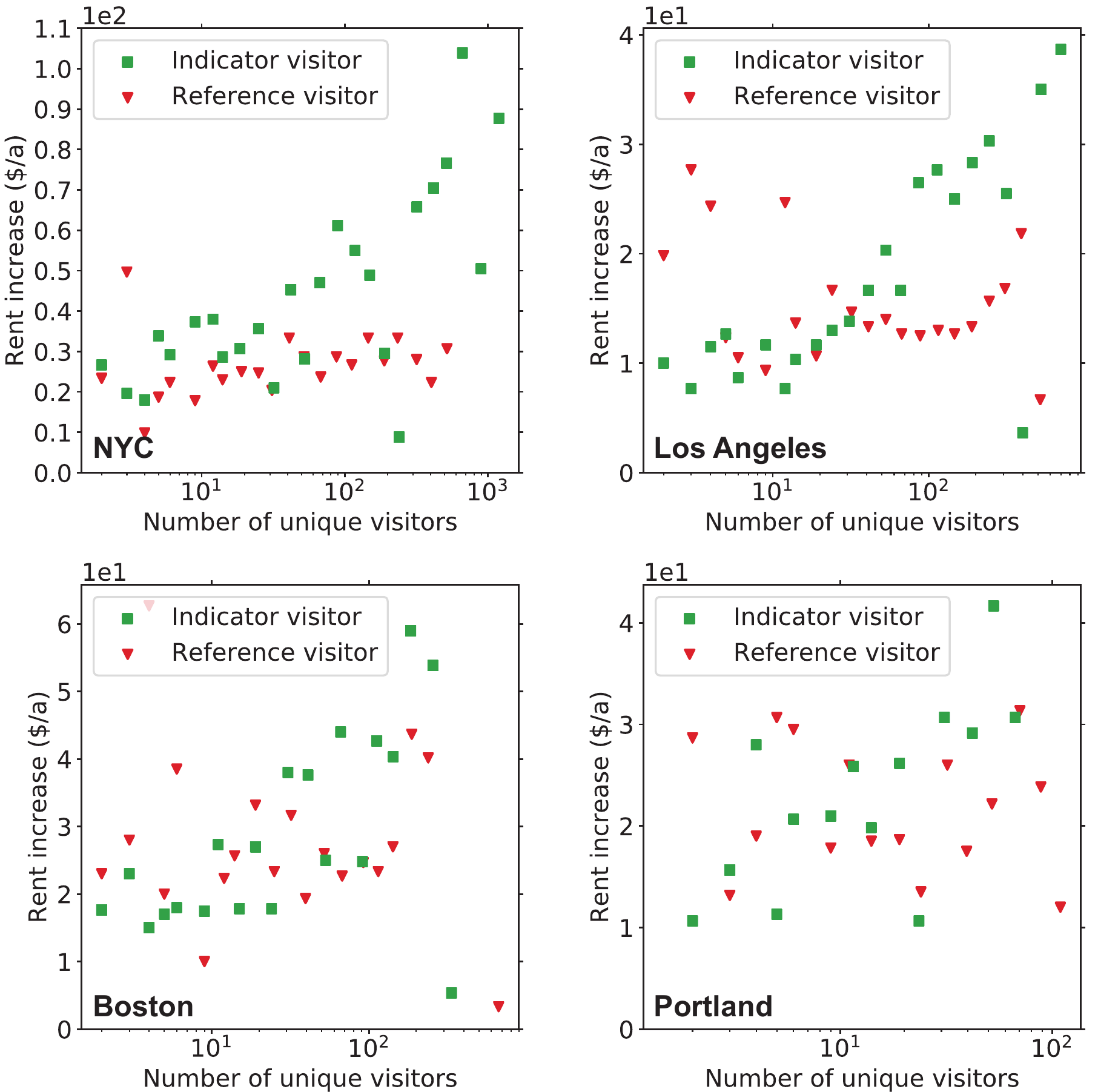}
    \caption{\csentence{Robustness of the indicator group across multiple cities.} Comparing the rent price predictions of the indicator group and the reference group for New York City (indicator group: $r_s=0.70$; $p<10^{-4}$; $N=25$, reference group: $r_s=0.46$; $p=0.03$; $N=22$), Los Angeles (indicator group: $r_s=0.73$; $p<10^{-4}$; $N=23$), Boston (indicator group: $r_s=0.57$; $p<10^{-2}$; $N=20$) and Portland (indicator group: $r_s=0.67$; $p<10^{-2}$; $N=14$). The correlation values for the reference group in Los Angeles, Boston and Portland are not significant ($p>0.05$).}
    \label{fig:deltarent_multiple}
\end{figure}
%
The robustness of our indicator group as a predictor of rising housing prices is confirmed for all cities analyzed in this paper, see Figure~\ref{fig:deltarent_multiple}.


\textbf{Transforming neighborhoods.} Neighborhoods that experience large increases in rental prices might already have a relatively high median rent. In simpler terms: an already expensive neighborhood becomes even more expense. Such rent increases might not always go hand-in-hand with larger changes in the characteristics of a neighborhood. On the other hand, neighborhoods that have relatively affordable rents but see similarly large increases in rent prices in subsequent years, could be more significantly affected, which is why this often holds the concern of urban planners and policy makers. However, finding and identifying these neighborhoods at an early stage is a challenging problem. This is partly caused by the delay in measuring these changes in conventional data sources. While social sensing might provide a solution, another challenge remains: real estate markets themselves might be slow to respond to neighborhood changes and rents could thus increase with a delay. Here we use this delay to our advantage and identify neighborhoods that attract the indicator group with the same frequency as other neighborhoods but have disproportionately low rents. More precisely, we define the Visitor-Rent-Index as the median rent of a neighborhood divided by the median rent of all the other neighborhoods in the same bin minus one. Negative VRI values thus indicate a disproportionately low rent, and vice versa.

\begin{figure}[t!]
    \includegraphics[width=0.9\linewidth]{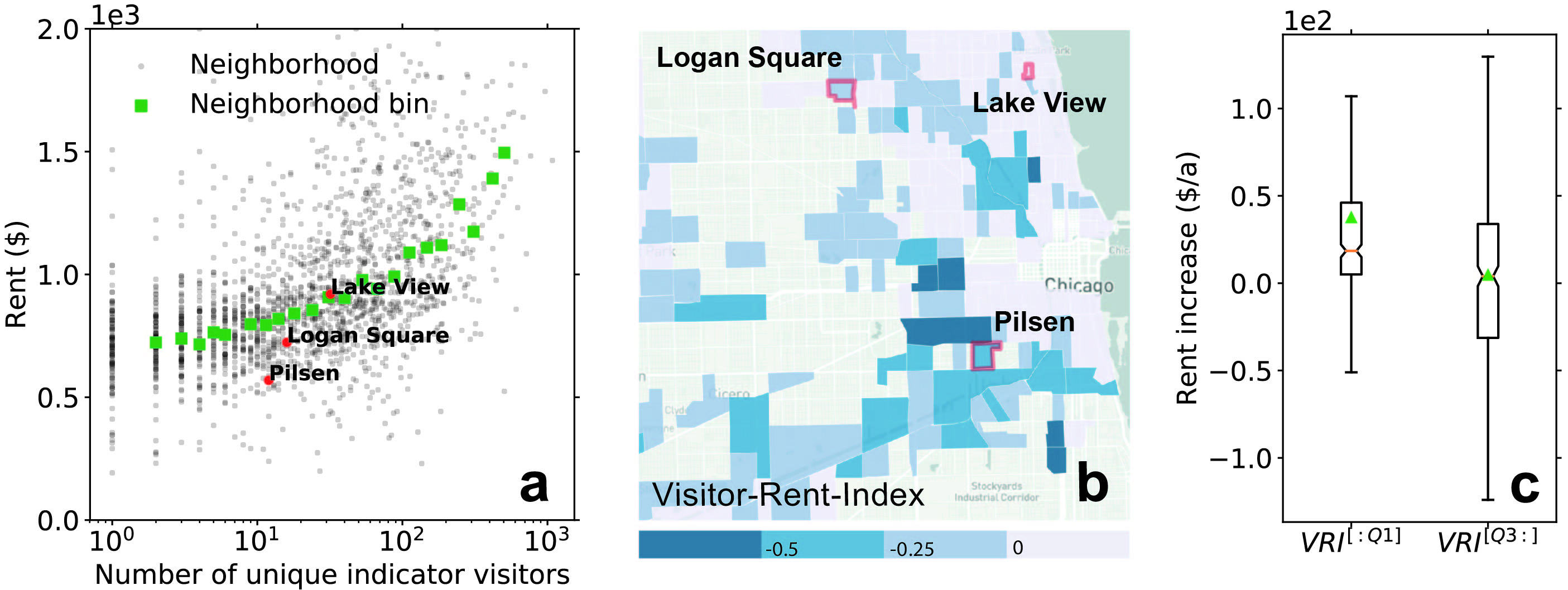}
    \caption{\csentence{Transforming neighborhoods.} \textbf{a}, For each neighborhood in Chicago, we compare the number of unique indicator visitors in 2012/2013 with the census median rent in 2012; the neighborhood bins (green) generate a baseline (corresponding to Figure \ref{fig:rent_Chicago}b). Neighborhoods below (above) the baseline show a disproportionately high (low) rent compare to other neighborhoods in the same bin. With Pilsen, Logan Square, and Lake View, we show three neighborhoods in different transformation stages. \textbf{b}, Mapping the level of disproportion with the Visitor-Rent-Index (VRI) (on the rise, $VRI < 0$). Here, we removed neighborhoods with less than 10 indicator visitors from the map. \textbf{c}, We compare the rent increase (2012-2015) in neighborhoods with a low VRI (first quartile) to neighborhoods with a high VRI (last quartile). Neighborhoods with a low VRI show significantly higher rent increase.}
    \label{fig:nexthot_Chicago}
\end{figure}

The result is shown in Figure \ref{fig:nexthot_Chicago}a. It highlights three neighborhoods in different transformation stages: Pilsen, which in 2012 still had relatively low median rent and could be labeled as in an early stage of transformation ($VRI\ll0$); Logan Square, which in 2012 had moderately higher rents and can be seen as in a medium stage of transformation ($VRI<0$); and finally Lake View, which is in a later stage of transformation with high rents ($VRI>=0$). At this point, we selected these well-known neighborhoods as illustration. Figure \ref{fig:nexthot_Chicago}b maps the Visitor-Rent-Index with \mbox{$VRI<0$} indicating neighborhoods in 2012 where future transformation is likely. Figure \ref{fig:nexthot_Chicago}c supports our hypothesis that indeed the neighborhoods with a low VRI show significantly higher rent increase in the following years compared to neighborhoods with a high VRI. However, more research needs to be done for a systematic analysis of VRI and its relation to neighborhood transformation.

It is here that the advantages of the combination of indicator groups with social sensing start to become clear. By utilizing the finer spatio-temporal granularity of data available from social media data platforms, we can potentially analyze and predict dynamic urban processes that are difficult to capture with conventional census or survey instruments, or can only be captured with significant delay. This potential has long been highlighted in the academic literature but so far it has been difficult to demonstrate the potential in more applied, practical work as such datasets have large issues with bias and representativeness. In addition, once we go down to those finer spatial and temporal scales, big data becomes `small' very fast. The indicator group approach may overcome such limitations because the overall size of an indicator group is not so relevant, as long as it is not too `rare' and is indeed a good indicator for the phenomenon of interest.


\textbf{Beyond income.} Up until now, income was the only user attribute included. However, the great potential of the indicator approach lies in its flexibility. First, we can include multiple user attributes to fine-tune the indicator group identification. For example, we could specify it to `high-income low-age single-status'. Second, we can include multiple indicator groups to improve the prediction power (corresponding to species assemblages in ecology). In this sense, selecting the appropriate indicator groups is not unlike other optimization problems. In future work, we can make use of machine learning techniques to further enhance the predictive power and accuracy of our method. This will help us for the inference of more detailed socio-economic characteristics of urban dwellers \cite{Blumenstock2015predicting} as well as for the systematic analysis of different indicator groups and their combinations.

For this work, we illustrate this potential by introducing a second user attribute. By way of example, we adopt the median age of a neighborhood. Importantly, age as a single attribute might not be a good predictor for rising housing prices (consider neighborhoods with a high density of low-age blue collar families). It is only in its combination with income that it increases prediction power where `high-income low-age' people can indicate `hipness' of a neighborhood. We evaluate the user's age profile in a similar way to its income profile. This time, whenever a user visited a neighborhood we assign the weighted median age instead of the median household income. We then upgrade the indicator group by only selecting the 10 percent `youngest' of the high-income-profile users. In the following we refer to them as high-income-low-age-profile users. Consequently, the upgraded indicator measure counts the number of unique high-income-low-age-profile users a neighborhood attracted between July 2012 and June 2013.

\begin{figure}[t!]
    \includegraphics[width=0.75\linewidth]{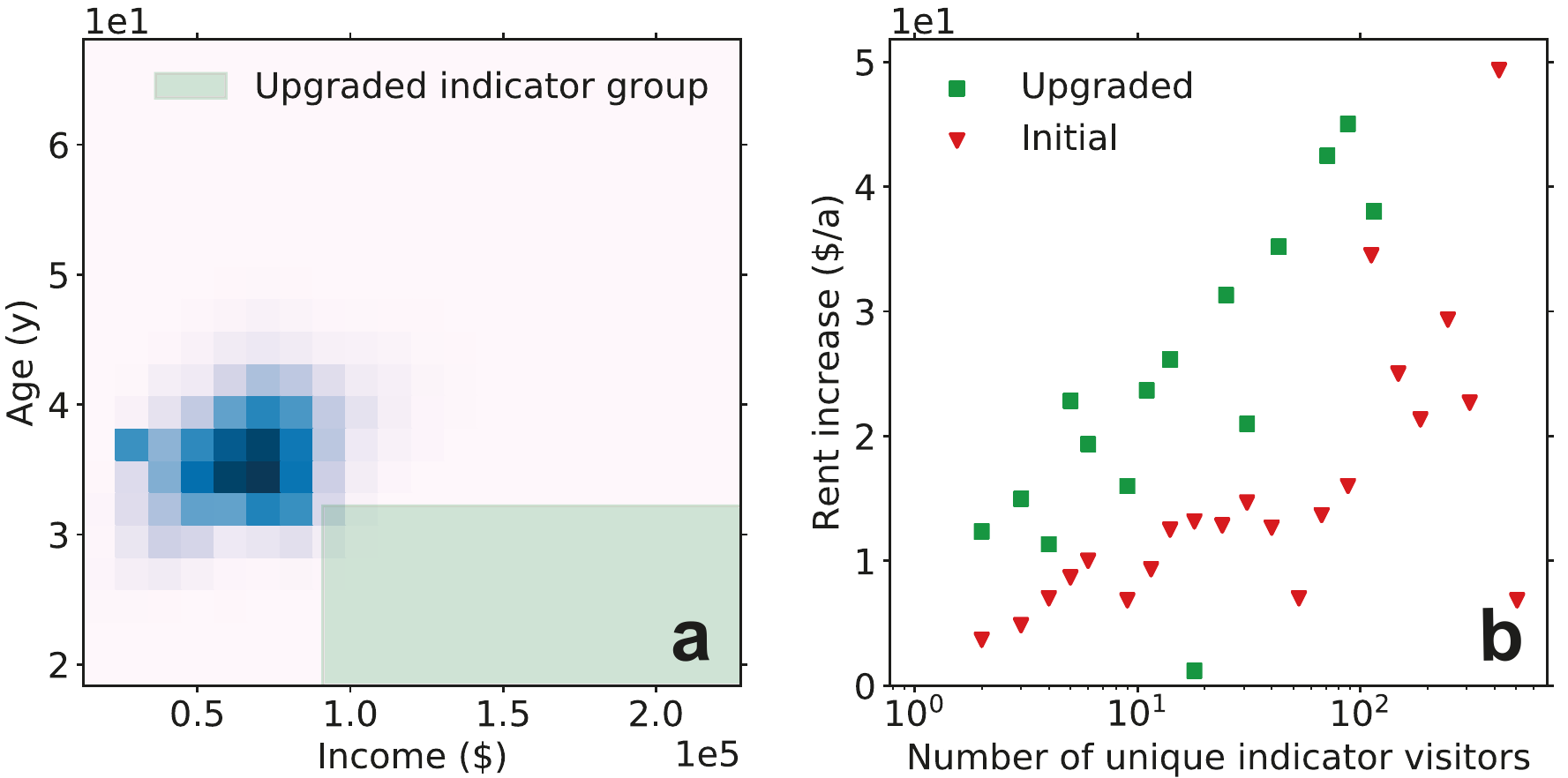}
    \caption{\csentence{Upgrading the indicator group.} \textbf{a}, Income-age-profile distribution for users in Chicago. We assign an income-age profile to each user based on the census median income and median age of the visited neighborhoods (darker colors indicate a higher frequency). The 10 percent `youngest' of the `richest' users represent the upgraded indicator group (green) while all `richest' users represent the initial indicator group. \textbf{b}, For each neighborhood bin in Chicago, we compare the number of upgraded indicator visitors (green) and the number of initial indicator visitors (red) in 2012/2013 with the 2012-2015 change of the census median rent. Using visitors with a high-income-low-age profile as indicator group yields a stronger correlation than using all high-income-profile visitors (upgraded: $r_s=0.78$; $p<10^{-3}$; $N=15$, initial: $r_s=0.70$; $p<10^{-3}$; $N=22$).}
    \label{fig:canary2_Chicago}
\end{figure}

Figure \ref{fig:canary2_Chicago}a shows the distribution of the income-age profile for Chicago users with the upgraded indicator group (green) now representing the 10 percent `youngest' of the `richest' users. Figure \ref{fig:canary2_Chicago}b shows that the number of high-income low-age visitors is a better indicator for high rents than high-income visitors in general, which confirms our initial assumption that age as a second user attribute can significantly improve the prediction power of the indicator group.

\section*{Conclusion}

This paper introduced the concept of indicator groups, adapted from ecology, to the analysis of socio-economic processes in cities. Specifically, it uses socio-economic profiles of LBSN users in combination with their activity spaces as predictor of changing urban neighborhoods. It does so by defining an indicator group as a small, specifically defined, subset of the population that is especially sensitive to the neighborhood changes that we are interested in studying. This is akin to the canaries that miners deployed in coal mines: canaries are more sensitive to carbon-monoxide than humans and thus serve a natural early warning system. Our approach has shown to be particularly relevant for the early detection of future drastic rent increases in neighborhoods that currently have relatively low rents. This is an area of study that is highly relevant to urban stakeholders, from policymakers to real estate developers. Traditional approaches can be hindered by a lack of granular data or conversely an overload of data (needle in a haystack). The indicator group approach provides a potential solution. If defined appropriately for the phenomenon of interest, this approach can yield early predictions while simultaneously reducing the amount of data that needs to be collected and analyzed.

Moreover, the indicator group approach is flexible in nature. Different groups can be identified for different social processes of interest, or they can even be adapted to unique local contexts. While we demonstrated the feasibility of our framework for the specific problem of predicting rising housing prices, it thus can be applied to a broader spectrum of urban processes. Possible examples include crime patterns or increasing social segregation in a city. Nevertheless, further investigation needs to be forthcoming to enable a more systematic identification of suitable indicator groups. With the increasing availability of detailed data on how individuals actually make use of urban space, finding a suitable indicator group can be seen as an optimization problem. On these premises, the application of machine learning techniques may offer a promising next step towards a novel urban monitoring tool that is based on a comprehensive set of early-warning \mbox{indicators \cite{scheffer2009early,carpenter2011early}}.

\begin{backmatter}



\section*{Competing interests}
The authors declare that they have no competing interests.

\section*{Funding}
This research was supported by the Singapore-ETH Centre, which was established collaboratively between ETH Zurich and Singapore's National Research Foundation (FI 370074016) under its Campus for Research Excellence and Technological Enterprise programme.

\section*{Author's contributions}
All authors designed the study, performed research, discussed the results and wrote the manuscript. A.A.S. and A.P. processed and analysed the data.

\section*{Acknowledgements}
The authors thank Lukas Lienhart for pointing us to the concept of bioindicators as applied in ecology. A.P. acknowledges the support of the University of Kentucky in providing the data collection infrastructure that enabled the use of the Twitter dataset in this paper.




\bibliographystyle{bmc-mathphys} 

\end{backmatter}
\end{document}